\documentclass[twocolumn,amsmath,amssymb,amsfonts,superscriptaddress,floatfix,showpacs,prd,aps,nofootinbib]{revtex4}
\usepackage[utf8x]{inputenc}
\usepackage{graphicx,color,mathrsfs,flafter}
\usepackage[dvipsnames]{xcolor}
\usepackage[colorlinks=true,urlcolor=Blue,linkcolor=Blue]{hyperref}
\usepackage[all]{hypcap}

\begin{document}
\title{Schottky mass measurements of heavy neutron-rich nuclides in the element range $70\leq Z\leq79$ at the ESR}

\author{D.~Shubina}
\affiliation{Max-Planck-Institut f\"ur Kernphysik, Saupfercheckweg 1, 69117 Heidelberg, Germany}
\affiliation{Fakult\"at f\"ur Physik und Astronomie, Universit\"at Heidelberg, Philosophenweg 12, 69120 Heidelberg, Germany}
\affiliation{GSI Helmholtzzentrum f\"ur Schwerionenforschung, Planckstra{\ss}e 1, 64291 Darmstadt, Germany}
\author{R.B.~Cakirli}\affiliation{Max-Planck-Institut f\"ur Kernphysik, Saupfercheckweg 1, 69117 Heidelberg, Germany}
\affiliation{Department of Physics, University of Istanbul, Istanbul, Turkey}
\author{Yu.A.~Litvinov}\affiliation{Max-Planck-Institut f\"ur Kernphysik, Saupfercheckweg 1, 69117 Heidelberg, Germany}
%\affiliation{Fakult\"at f\"ur Physik und Astronomie, Universit\"at Heidelberg, Philosophenweg 12, 69120 Heidelberg, Germany}
%\affiliation{Physikalisches Institut, Universit\"at Heidelberg, Philosophenweg 12, 69120 Heidelberg, Germany}
\affiliation{GSI Helmholtzzentrum f\"ur Schwerionenforschung, Planckstra{\ss}e 1, 64291 Darmstadt, Germany}
\author{K.~Blaum}\affiliation{Max-Planck-Institut f\"ur Kernphysik, Saupfercheckweg 1, 69117 Heidelberg, Germany}
%\affiliation{Physikalisches Institut, Universit\"at Heidelberg, Philosophenweg 12, 69120 Heidelberg, Germany}
\author{C.~Brandau}\affiliation{GSI Helmholtzzentrum f\"ur Schwerionenforschung, Planckstra{\ss}e 1, 64291 Darmstadt, Germany}
\affiliation{ExtreMe Matter Institute EMMI, GSI Helmholtzzentrum f\"ur Schwerionenforschung, 64291 Darmstadt, Germany}
\author{F.~Bosch}\affiliation{GSI Helmholtzzentrum f\"ur Schwerionenforschung, Planckstra{\ss}e 1, 64291 Darmstadt, Germany}
\author{J.J.~Carroll}\affiliation{US Army Research Laboratory, 2800 Powder Mill Road, Adelphi MD, USA}
\author{R.F.~Casten}\affiliation{Wright Nuclear Structure Laboratory, Yale University, New Haven, Connecticut 06520, USA}
\author{D.M.~Cullen}\affiliation{Schuster Laboratory, University of Manchester, Manchester M13 9PL, United Kingdom}
\author{I.J.~Cullen}\affiliation{Department of Physics, University of Surrey, Guildford, Surrey GU2 7XH, United Kingdom}
\author{A.Y.~Deo}\affiliation{Department of Physics, University of Surrey, Guildford, Surrey GU2 7XH, United Kingdom}
\author{B.~Detwiler}\affiliation{Youngstown State University, One University Plaza, Youngstown, Ohio 44555, USA}
\author{C.~Dimopoulou}\affiliation{GSI Helmholtzzentrum f\"ur Schwerionenforschung, Planckstra{\ss}e 1, 64291 Darmstadt, Germany}
\author{F.~Farinon}\affiliation{GSI Helmholtzzentrum f\"ur Schwerionenforschung, Planckstra{\ss}e 1, 64291 Darmstadt, Germany}
\author{H.~Geissel}\affiliation{GSI Helmholtzzentrum f\"ur Schwerionenforschung, Planckstra{\ss}e 1, 64291 Darmstadt, Germany}
\affiliation{II Physikalisches Institut, Justus-Liebig-Universit\"at Gie{\ss}en, 35392 Gie{\ss}en, Germany}
\author{E.~Haettner}\affiliation{II Physikalisches Institut, Justus-Liebig-Universit\"at Gie{\ss}en, 35392 Gie{\ss}en, Germany}
\author{M.~Heil}\affiliation{GSI Helmholtzzentrum f\"ur Schwerionenforschung, Planckstra{\ss}e 1, 64291 Darmstadt, Germany}
\author{R.S.~Kempley}\affiliation{Department of Physics, University of Surrey, Guildford, Surrey GU2 7XH, United Kingdom}
\author{C.~Kozhuharov}\affiliation{GSI Helmholtzzentrum f\"ur Schwerionenforschung, Planckstra{\ss}e 1, 64291 Darmstadt, Germany}
\author{R.~Kn\"obel}\affiliation{GSI Helmholtzzentrum f\"ur Schwerionenforschung, Planckstra{\ss}e 1, 64291 Darmstadt, Germany}
\author{J.~Kurcewicz}\affiliation{GSI Helmholtzzentrum f\"ur Schwerionenforschung, Planckstra{\ss}e 1, 64291 Darmstadt, Germany}
\author{N.~Kuzminchuk}\affiliation{II Physikalisches Institut, Justus-Liebig-Universit\"at Gie{\ss}en, 35392 Gie{\ss}en, Germany}
\author{S.A.~Litvinov}\affiliation{GSI Helmholtzzentrum f\"ur Schwerionenforschung, Planckstra{\ss}e 1, 64291 Darmstadt, Germany}
\author{Z.~Liu}\affiliation{School of Physics and Astronomy, University of Edinburgh, Edinburgh EH9 3JZ, United Kingdom}
\author{R.~Mao}\affiliation{Institute of Modern Physics, Chinese Academy of Sciences, Lanzhou 730000, People's Republic of China}
\author{C.~Nociforo}\affiliation{GSI Helmholtzzentrum f\"ur Schwerionenforschung, Planckstra{\ss}e 1, 64291 Darmstadt, Germany}
\author{F.~Nolden}\affiliation{GSI Helmholtzzentrum f\"ur Schwerionenforschung, Planckstra{\ss}e 1, 64291 Darmstadt, Germany}
\author{Z.~Patyk}\affiliation{National Centre for Nuclear Research, PL-00681 Warsaw, Poland}
\author{W.R.~Plass}\affiliation{II Physikalisches Institut, Justus-Liebig-Universit\"at Gie{\ss}en, 35392 Gie{\ss}en, Germany}
\author{A.~Prochazka}\affiliation{GSI Helmholtzzentrum f\"ur Schwerionenforschung, Planckstra{\ss}e 1, 64291 Darmstadt, Germany}
\author{M.W.~Reed}\affiliation{Department of Physics, University of Surrey, Guildford, Surrey GU2 7XH, United Kingdom}
\affiliation{Department of Nuclear Physics, R.S.P.E., Australian National University, Canberra ACT 0200, Australia}
\author{M.S.~Sanjari}\affiliation{GSI Helmholtzzentrum f\"ur Schwerionenforschung, Planckstra{\ss}e 1, 64291 Darmstadt, Germany}
\affiliation{Goethe-Universit\"at Frankfurt, 60438 Frankfurt, Germany}
\author{C.~Scheidenberger}\affiliation{GSI Helmholtzzentrum f\"ur Schwerionenforschung, Planckstra{\ss}e 1, 64291 Darmstadt, Germany}
\affiliation{II Physikalisches Institut, Justus-Liebig-Universit\"at Gie{\ss}en, 35392 Gie{\ss}en, Germany}
\author{M.~Steck}\affiliation{GSI Helmholtzzentrum f\"ur Schwerionenforschung, Planckstra{\ss}e 1, 64291 Darmstadt, Germany}
\author{Th.~St\"ohlker}\affiliation{GSI Helmholtzzentrum f\"ur Schwerionenforschung, Planckstra{\ss}e 1, 64291 Darmstadt, Germany}
\affiliation{Friedrich-Schiller-Universit\"at Jena, 07737 Jena, Germany}
\affiliation{Helmholtz-Institut Jena, 07743 Jena, Germany}
\author{B.~Sun}\affiliation{II Physikalisches Institut, Justus-Liebig-Universit\"at Gie{\ss}en, 35392 Gie{\ss}en, Germany}
%\affiliation{GSI Helmholtzzentrum f\"ur Schwerionenforschung, Planckstra{\ss}e 1, 64291 Darmstadt, Germany}
\affiliation{School of Physics and Nuclear Energy Engineering, Beihang University, 100191 Beijing, PRC}
\author{T.P.D.~Swan}\affiliation{Department of Physics, University of Surrey, Guildford, Surrey GU2 7XH, United Kingdom}
\author{G.~Trees}\affiliation{Youngstown State University, One University Plaza, Youngstown, Ohio 44555, USA}
\author{P.M.~Walker}\affiliation{Department of Physics, University of Surrey, Guildford, Surrey GU2 7XH, United Kingdom}
\affiliation{CERN, CH-1211 Geneva 23, Switzerland}
\author{H.~Weick}\affiliation{GSI Helmholtzzentrum f\"ur Schwerionenforschung, Planckstra{\ss}e 1, 64291 Darmstadt, Germany}
\author{N.~Winckler}\affiliation{Max-Planck-Institut f\"ur Kernphysik, Saupfercheckweg 1, 69117 Heidelberg, Germany}
\affiliation{GSI Helmholtzzentrum f\"ur Schwerionenforschung, Planckstra{\ss}e 1, 64291 Darmstadt, Germany}
\author{M.~Winkler}\affiliation{GSI Helmholtzzentrum f\"ur Schwerionenforschung, Planckstra{\ss}e 1, 64291 Darmstadt, Germany}
\author{P.J.~Woods}\affiliation{School of Physics and Astronomy, University of Edinburgh, Edinburgh EH9 3JZ, United Kingdom}
\author{T.~Yamaguchi}\affiliation{Graduate School of Science and Engineering, Saitama University, Saitama 338-8570, Japan}
\author{C.~Zhou}\affiliation{Wright Nuclear Structure Laboratory, Yale University, New Haven, Connecticut 06520, USA}

\begin{abstract}
Storage-ring mass spectrometry was applied to neutron-rich
$^{197}$Au projectile fragments. Masses of $^{181,183}$Lu,
$^{185,186}$Hf, $^{187,188}$Ta, $^{191}$W, and $^{192,193}$Re
nuclei were measured for the first time. The uncertainty of
previously known masses of $^{189,190}$W and $^{195}$Os nuclei was
improved. Observed irregularities on the smooth two-neutron
separation energies for Hf and W isotopes are linked to the
collectivity phenomena in the corresponding nuclei.
\end{abstract}
\maketitle
\section{Introduction}

The binding energy of the nucleus, or its mass, contains
information about interactions between its constituent protons and
neutrons. Precision mass data as well as separation energies
extracted from them can unveil much about the underlying nuclear
structure~\cite{Blaum,Blaum2}. For instance, magic numbers can easily be seen from
separation energies which reveal a huge drop (depending on the
size of the closed shell), or a jump or a gap (depending on the way the
separation energy is constructed) after a magic number~\cite{Novikov}. 
In addition, separation energies can reflect
collective effects as we will discuss below. Such data are
therefore valuable in understanding the underlying shell structure
in nuclei and the evolution of collective effects, and are
therefore essential to provide the basis for the development and
testing of a comprehensive theory of nuclei. In turn, a reliable
nuclear theory is of utmost importance for calculating the
properties of unknown nuclei, which are needed, for example, in
the modelling of the rapid neutron capture process (r-process) of
nucleosynthesis in stars~\cite{Bertulani}. 
%The nuclei involved in this process lie
%far away from the valley of $\beta$-stability and are barely
%within reach of the present radioactive beam facilities.
%Furthermore, even if the neutron-rich nuclei at the N = 50 and N =
%82 shell closures were to  become accessible at future facilities
%like FAIR, FRIB, SPIRAL2 etc. (see, e.g., Ref.~\cite{Bertulani}),
%many of the nuclei at the $N=126$ shell closure, which are
%believed to be responsible for the elemental abundance peak at
%$A\sim195$, will remain inaccessible. 
Therefore, new data on neutron-rich heavy
nuclei are essential.

However, such data are scarce mainly due to the complexity of
producing these exotic nuclei. This becomes evident if one
glances at the chart of nuclides (see e.g. Ref.~\cite{nchart}) where
for elements $Z\sim70-80$ the number of observed neutron-rich
nuclides is very limited. Very recently, several tens of new
isotopes were discovered in projectile fragmentation of uranium
beams, though no spectroscopic information could yet be obtained for
these nuclei~\cite{kurcewicz}. Their production rates are tiny, 
which requires very efficient measurement
techniques. One such technique for mass measurements is
storage-ring mass spectrometry~\cite{FGM}.

In this paper we report on direct mass measurements of
neutron-rich nuclei in the element range from lutetium to osmium.
Masses for nine nuclei were measured for the first time, and for
three nuclei the mass uncertainty was improved. 
It is known that nuclear collective effects can be seen in the behavior of nucleon separation energies~\cite{Cakirli2}.
Here, we investigate the relation between rather subtle effects in two-neutron separation energies, $S_{2n}$,
and changes in both collectivity and neutron number.
Observed irregularities in the smooth two-neutron separation
energies for Hf and W isotopes are linked to changes in
collective observables. The importance of the number of valence
nucleons is discussed in the context of collective
contributions to binding energies calculated with the IBA
model~\cite{IBA}.
%As mentioned above, collective effects can be seen from separation energies~\cite{Cakirli2}. For example, in a well-known shape
%transitional region, $Z\sim64$, $N\sim90$, sudden changes in structure can be seen from $S_{2n}$ values as a function of
%neutron number. In this region, the structure, for example, in Gd from $N=88$ to 90, changes suddenly from spherical to deformed.
%Beyond $N=90$, the known nuclei are well deformed. This change from spherical to deformed shows up as a flattening in $S_{2n}$
%versus $N$ just after $N=90$ (relative to its downward trend) due to a gain in binding energy as nuclei become more deformed. The
%new data in this work show that changes in the trends of $S_{2n}$ values can occur for other reasons as well. Here, we will
%investigate the relation between rather subtle effects in $S_{2n}$, revealed by the present data, and changes in collectivity, and in the numbers of valence nucleons.

\section{Experiment}
%%%%%%%%%%%%%%%%
\begin{figure*}[t!]
\centering
\includegraphics[width=\textwidth]{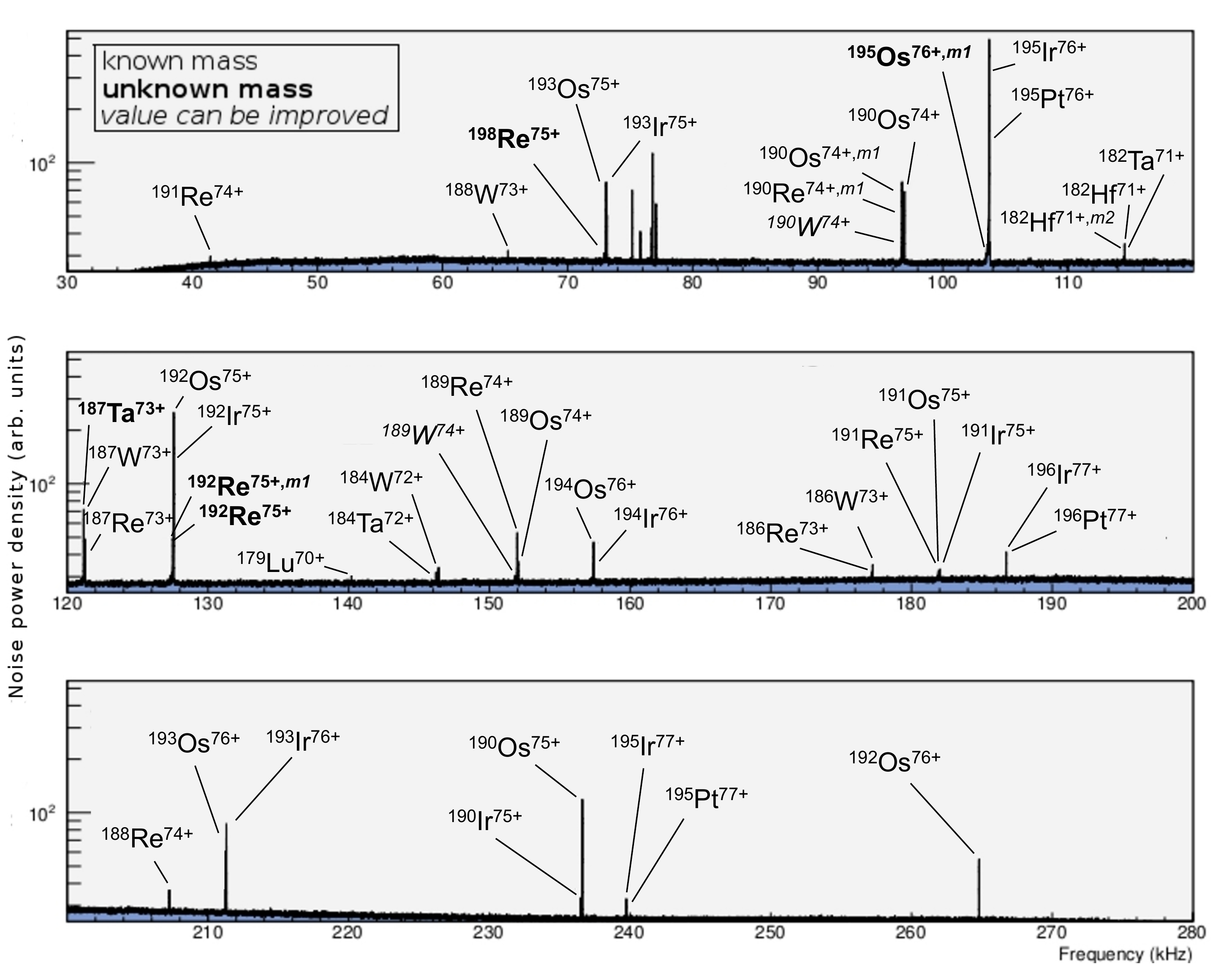}
\caption{Example of a Schottky frequency spectrum with the
corresponding isotope identification. This spectrum is
combined from eight independent injections acquired for the electron
cooler voltage $U_c=209$~kV. Nuclides with known and previously
unknown masses are indicated by different fonts (see legend).}
\label{id}
\end{figure*}
%%%%%%%%%%%%%%%%
\begin{figure}[b]
\centering
\includegraphics[width=\linewidth]{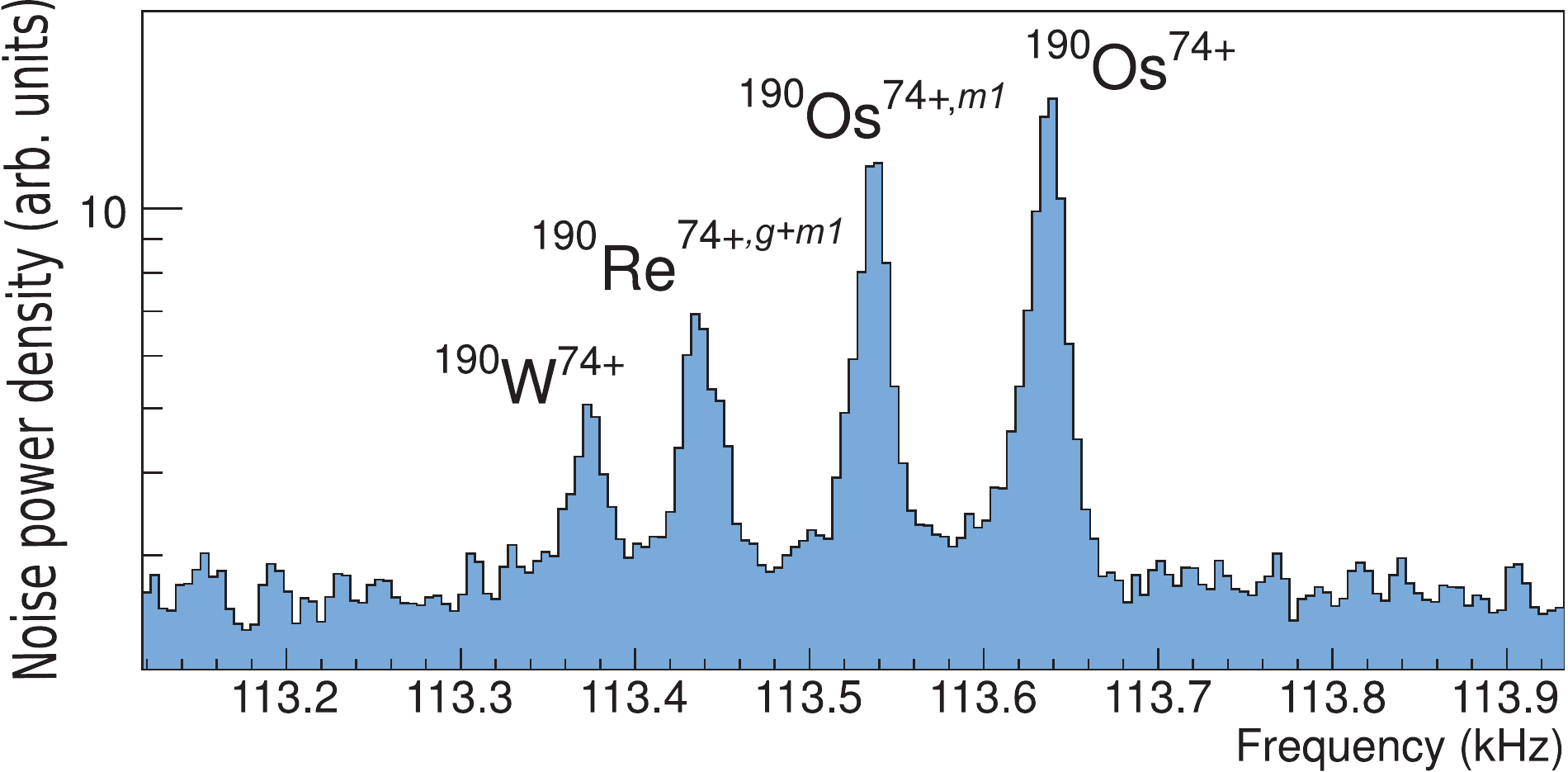}
\caption{A zoom of the Schottky frequency spectrum illustrated in
Fig.~\ref{id} on a quadruplet of $A=190$ isobars. The peaks
corresponding to nuclides in isomeric states are labeled with
$m$.} \label{spzoom}
\end{figure}
%%%%%%%%%%%%%%%%
\begin{figure*}[t]
\centering
\includegraphics[width=\linewidth]{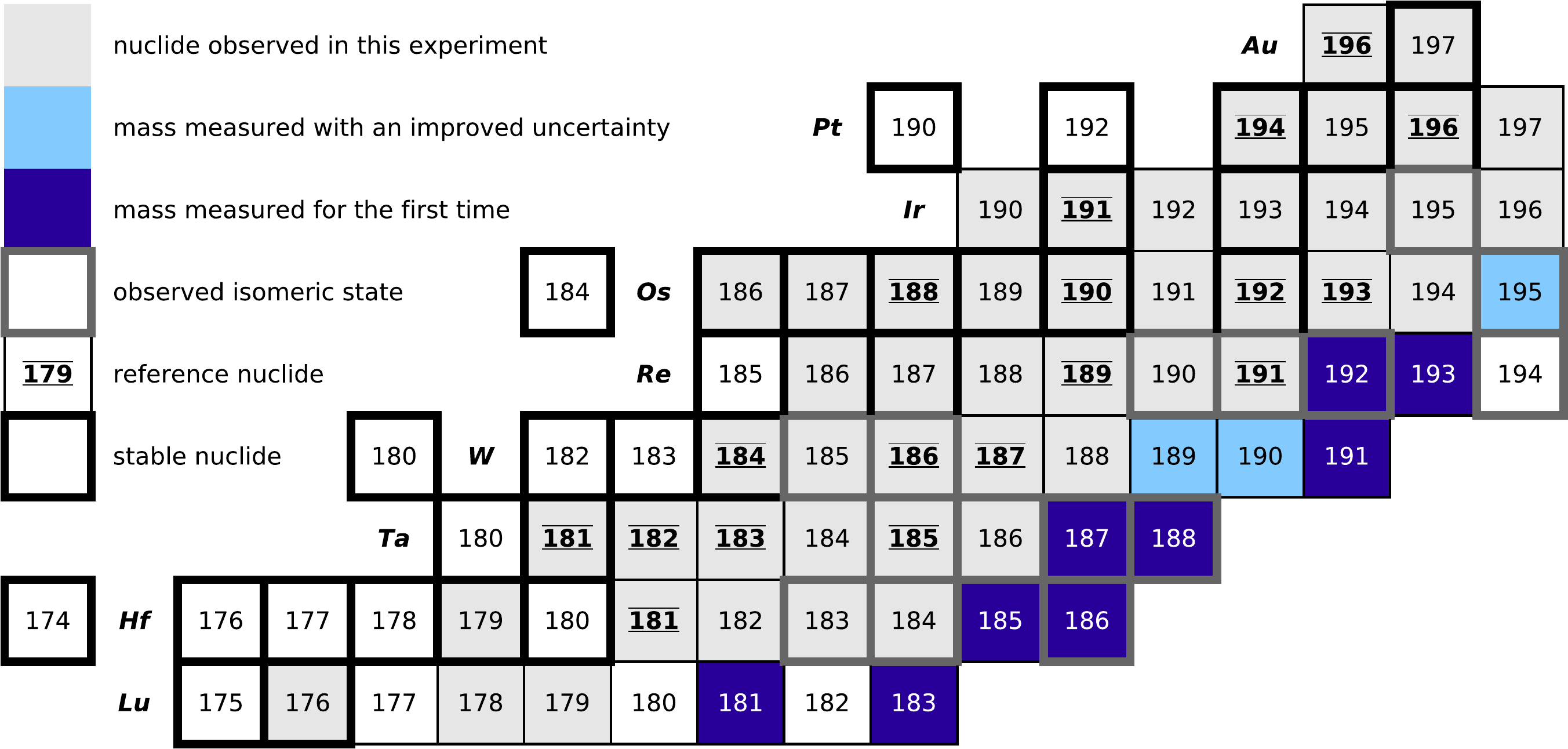}
\caption{(Color online) A part of the chart of nuclides indicating the
nuclides measured in this work as well as the nuclides in the
ground and isomeric states identified in the other part of this
experiment devoted to the search for new K-isomers in this
region~\cite{Matt,Matt2}.} \label{nchart}
\end{figure*}
%%%%%%%%%%%%%%%%
The experiment was conducted at GSI Helmholtzzentrum f\"ur
Schwerionenforschung in Darmstadt. Heavy neutron-rich nuclei of
interest were produced in projectile fragmentation of $^{197}$Au
primary beams. The experiment described here was part of a
larger experimental campaign, some results of which are described
in Refs.~\cite{Matt,Matt2,Matt25,Matt3}. The $^{197}$Au beams were accelerated
to the energy of 11.4~MeV/u in the linear accelerator UNILAC and
then injected into the heavy ion synchrotron SIS-18~\cite{sis},
where they were further accelerated to an energy of 469.35~MeV/u.
The $^{197}$Au$^{65+}$ beams were fast extracted (within about
1~$\mu$s) and focused on a production target located at the
entrance of the fragment separator FRS~\cite{frs,Geissel1992NIM}.
As target we used 1036~mg/cm$^2$ thick $^9$Be with a 221~mg/cm$^2$
Nb backing for more efficient electron stripping. The reaction
products emerged from the target as highly-charged ions having
mostly 0, 1, or 2 bound electrons. The nuclides of interest were
transported through the FRS, being operated as a pure magnetic
rigidity ($B\rho$) analyzer~\cite{Geissel1992NIM}, and injected
into the cooler-storage ring ESR~\cite{esr}. The transmission
through the FRS and the injection into the ESR were optimized with
the primary beam, and the magnetic setting of FRS-ESR was fixed at
$B\rho=7.9$~Tm throughout the entire experiment. All ion species
within the acceptance of the FRS-ESR of about $\pm0.2$\% were
injected and stored. Only $25$\% of the ESR acceptance is filled
at the injection. We note, that in contrast to the settings
described in Refs.~\cite{Matt,Matt2}, in this experiment no
energy-loss degraders were employed in the FRS.

The relationship between relative revolution frequencies ($f$),
relative mass-over-charge ratios ($m/q$) and velocities ($v$) of
the particles stored in a ring is given
by~\cite{FGM,Ra-PRL,Ra-NPA,LiBo}:
\begin{equation}
\label{sms1}
\frac{\Delta f}{f}=-\alpha_p\frac{\Delta \frac{m}{q}}{\frac{m}{q}}+(1-\alpha_p\gamma^2)\frac{\Delta v}{v},
\end{equation}
where $\gamma$ is the relativistic Lorentz factor, $\alpha_p$
is the momentum compaction factor, which characterizes
the relative variation of the orbit length of stored particles per
relative variation of their magnetic rigidity (for more details
see Refs.~\cite{FGM,Ra-PRL,Ra-NPA,LiBo}). For the ESR, $\alpha_p$
is nearly constant for the entire revolution frequency acceptance
and is $\alpha_p\approx0.179$. From Eq.~\eqref{sms1} it becomes
obvious that the revolution frequency is a measure of the
mass-over-charge ratios of the stored ions provided that the
second term on the right hand side, containing the velocity spread
($\Delta v/v$), can be eliminated. The latter is achieved by
applying electron cooling~\cite{elcool}. For this purpose the
stored ions are merged over a length of about 2.5~m with a
continuous beam of electrons in the electron cooler device. The
mean circumference of the ESR is 108.4~m and at our energies the
ions circulate with a frequency of about 2~MHz passing the
electron cooler at each revolution. The energy of the electrons is
very accurately defined by the applied acceleration potential.
Within a few seconds the mean velocity of the ions becomes equal
to the mean velocity of the electrons. The velocity spread of the
stored ions, which is $\Delta v/v\approx4\cdot10^{-3}$ at the injection,
is thereby reduced to $\Delta v/v\approx10^{-7}$~\cite{elcool}.

The Schottky mass spectrometry (SMS) technique has been applied to
the electron cooled ions~\cite{Ra-NPA,FGM}. In this technique,
every stored highly-charged ion at each revolution in the ESR
induces mirror charges on a couple of parallel electrostatic
copper plates, the Schottky pick-up installed inside the ring aperture.
The noise from the pick-up, which is dominated by the
thermal noise, is amplified by a broad-band
low-noise amplifier~\cite{Schaaf}. In the present experiment, we analyzed the
noise power at about 60~MHz, corresponding to the 30$^{th}$
harmonic of the revolution frequency of the stored ions. The
pick-up signal was down-mixed using a $\sim$60~MHz reference
frequency from an external frequency generator. The acceptance of
the ESR at the 30$^{th}$ harmonic corresponds to about
320~kHz~\cite{Li-2004,Li-NPA}. Therefore, to cover the entire ESR
acceptance we digitized the signal with a sampling frequency of
640~kHz using a commercial 16-bit ADC~\cite{Kaza}. A Fourier
transform of the digitized data yielded the noise-power density
spectrum, or the Schottky frequency spectrum~\cite{FGM,Li-NPA}.

%In the present experiment we injected new ions every few minutes.
New ions were injected every few minutes. At the injection into
the ESR, the previously stored ions were removed. Several Schottky
frequency spectra were created for each injection. The parameters
of the Fourier transform algorithm were optimized offline. A
frequency resolution of 4.77~Hz/channel was chosen, which
corresponds to the time resolution of 0.21~s per spectrum.
Furthermore, every 50 consecutive Schottky spectra were averaged
to enhance the signal-to-noise ratio. Thus, Schottky spectra
integrated over 10~s were produced. The latter means that several
independent subsequent frequency spectra were obtained for each
injection of the ions into the ESR.

The electron cooling forces the ions to the same mean velocity
thus filling the entire acceptance of the ESR of $\Delta
B\rho/B\rho\sim\pm1.5\%$ (see Ref.~\cite{Ra-NPA}). Since $B\rho=m
v \gamma / q$, by changing the velocity of the electrons in
different injections, ions with different $m/q$ can be studied. In
the present experiment we varied the electron cooler voltage in
the range from 204~kV to 218~kV. On average eight injections were
recorded for each cooler setting. In order to facilitate the
assignment of the revolution frequencies with the corresponding
isotope identification, all spectra within each cooler setting
were combined together. In this case the maximum number of ion
species present in each setting can be used for the
identification. The latter is done based on Eq.~\eqref{sms1}.
As a starting point for the identification we used
the frequency of the stored $^{197}$Au$^{76+}$ primary ions. An
example of the combined Schottky frequency spectrum for an electron
cooler voltage of $U_c=209$~kV is illustrated in Fig.~\ref{id}.
Fig.~\ref{spzoom} shows a zoom on a quadruplet of lines of
$A=190$ isobars present in ground and/or isomeric states. The
latter are indicated with a label $m$. The peak finding and the
isotope identification were done automatically with a dedicated
ROOT-based~\cite{ROOT} software~\cite{Dasha}. The nuclides
observed in this experiment are illustrated on the chart of nuclides
in Fig.~\ref{nchart} together with the nuclides in the ground
and isomeric states identified in the other part of this
experiment (see Refs.~\cite{Matt,Matt2,Matt25,Matt3}).

\section{Data Analysis and Results}
\begin{table}[t!]
%\label{references}
\caption{Nuclides with accurately known masses used as references
to calibrate Schottky frequency spectra. Listed are the proton
($Z$) and mass ($A$) numbers, the number of experimental settings 
($N_{set}$) in which this reference mass was observed, literature
mass excess values from the Atomic-Mass Evaluation ~\cite{AME} 
($ME_{AME}$) as well as the re-determined mass excess values
($ME$) (see text) with the corresponding $\sigma_{stat}$ uncertainty 
%($\sqrt{\sigma^2_{stat}+\sigma^2_{syst}}$) uncertainty 
and its difference to the literature value
($\delta=ME-ME_{AME}$). 
Note that the systematic uncertainty of $\sigma_{syst}=38$~keV (see text) is 
not added here.
\label{references}}
\begin{center}
\begin{tabular}{cccccc}
\hline
\hline
Z & A &  $N_{set}$ & $ME_{\rm AME}$ & $ME$& $\delta$ \\
   &     &                     &  (keV)                      & (keV)                      &         (keV)     \\
\hline
72 & 181 & 1 & -47412(2) & -47412(40) & 0(40)  \\
\hline
73 & 181 & 1 & -48442(2) & -48383(40)& 59(40)  \\
73 & 182 & 3 & -46433(2) & -46466(29)& -32(29)  \\
73 & 183 & 3 & -45296(2) & -45276(16)& 20(16)  \\
73 & 185 & 6 & -41396(14)& -41350(14)& 46(20)  \\
\hline
74 & 184 & 5 & -45707(1) & -45663(17)& 44(17)  \\
74 & 186 & 7 & -42510(2) & -42493(12)& 17(13)  \\
74 & 187 & 8 & -39905(2) & -39863(8)& 41(8)  \\
\hline
75 & 189 & 9 & -37978(8) & -38063(10)& -85(13)  \\
75 & 191 & 9 & -34349(10)& -34364(3)& -15(11)  \\
\hline
76 & 188 & 7 & -41136(1) & -41115(12)& 21(12)  \\
76 & 190 & 7 & -38706(2) & -38637(15)& 69(15)  \\
76 & 192 & 7 & -35881(3) & -35833(8)& 48(8)  \\
76 & 193 & 7 & -33393(3) & -33329(8)& 63(8)  \\
\hline
77 & 191 & 1 & -36706(2) & -36650(71)& 56(71) \\
\hline
78 & 194 & 4 & -34763(1) & -34779(24)& -16(24)  \\
78 & 196 & 9 & -32647(1) & -32655(4)& -7(4)  \\
\hline
79 & 196 & 5 & -31140(3) & -31126(4)& 14(5)  \\
\hline
\hline
\end{tabular}\end{center}
\end{table}

In order to determine the unknown mass-over-charge ratios, the
Schottky frequency spectra have to be
calibrated~\cite{Ra-NPA,Li-NPA}. For this purpose we selected the
nuclides which were identified in our spectra and for which masses are
known experimentally according to the Atomic-Mass Evaluation 2003
(AME)~\cite{AME}. We note that an update of the AME was made
available in 2011~\cite{AME11}, which however contains no new
information in the mass region studied here. 
The data of the present work were already included in the latest AME published very recently~\cite{AME12}.
Furthermore we
required that the reference masses were obtained by more than one
independent measurement technique and that there must exist no
other ionic species with a close mass-to-charge ratio which could simultaneously be stored in the ESR.
Also the peaks corresponding to long-lived isomeric states (we observed
$^{182}$Hf$^{\rm m1}$, $^{186}$W$^{\rm m2}$, $^{190}$Os$^{\rm m1}$ and $^{190}$Ir$^{\rm m2}$ isomers)
were not used for calibration. The list of reference masses is
given in Table~\ref{references}.

\begin{figure}[b]
\centering
\includegraphics[width=\linewidth]{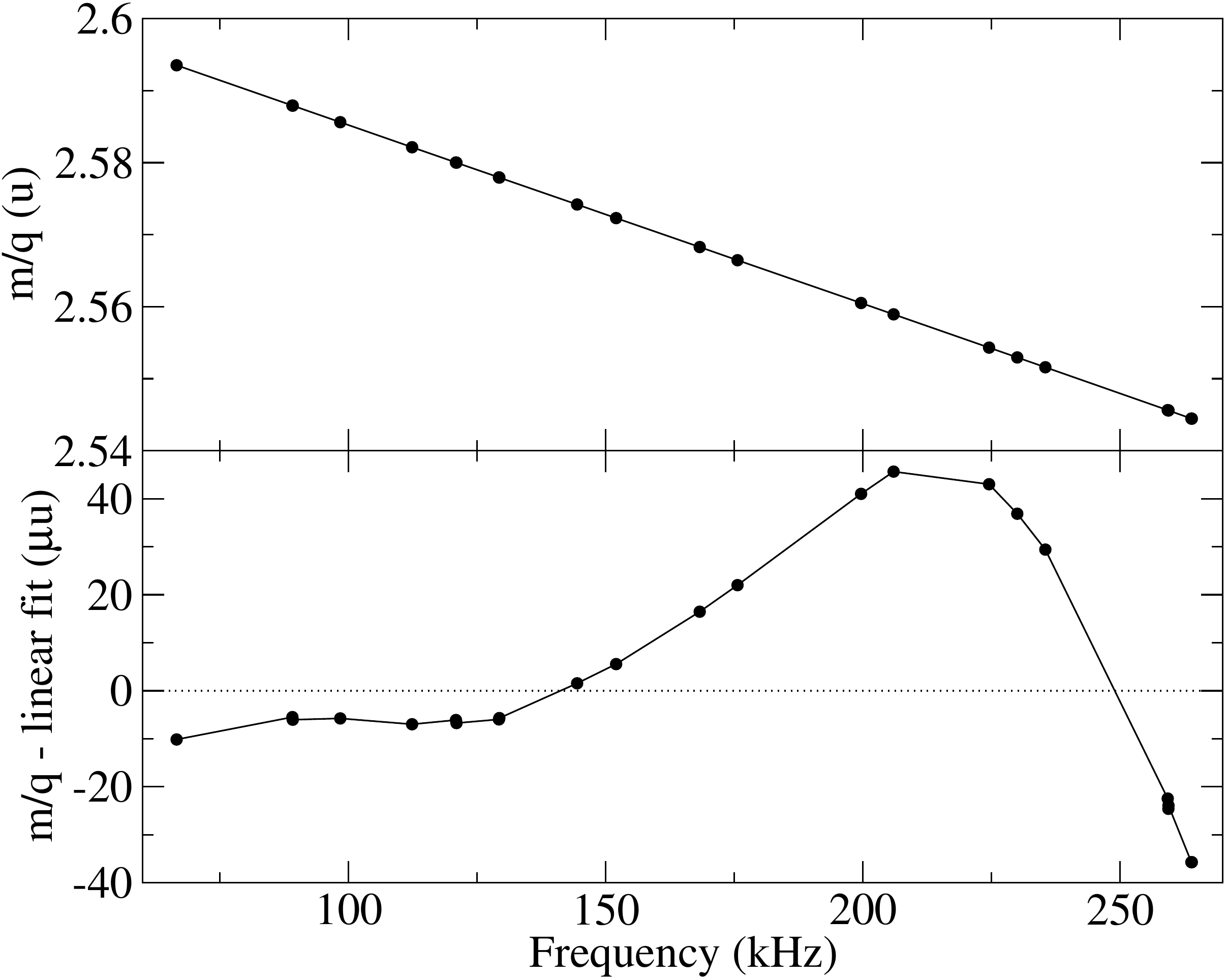} 
\caption{Top: Mass-over-charge ratio $m/q$ as a function of revolution
frequency. The solid line illustrates a straight line fit through
the calibration $m/q$-values. Bottom: The same as top but with
the subtracted linear fit.} \label{mqf}
\end{figure}

The momentum compaction factor $\alpha_p$, although nearly
constant, is a complicated function of the revolution frequency
$f$. If $\alpha_p$ were exactly constant, then the $m/q$ would linearly
depend on $f$. Fig.~\ref{mqf} (top) shows an example of a linear
fit through the calibration $m/q$-values for one of the measured spectra. The
residuals of the fit (bottom panel of the same figure) clearly
show that the calibration function is more complicated than a
low-order polynomial function. Polynomials of up to 4th order (5
free coefficients) were employed, but different compared to the analyses
performed in Refs.~\cite{Ra-NPA, Li-NPA,Chen2012}, the quality of
the fits was found unacceptable. 
A possible reason for the latter is the small number of reference masses in individual 10-s Schottky spectra.
Furthermore, due to time variations of the storage ring and electron cooler parameters, 
such as, e.g., their magnetic fields, it is not possible to establish a universal calibration curve. 
Therefore, we employed an analysis procedure 
in which we used linear splines to approximate the calibration curve in each individual spectrum. 
%Therefore, in the present analysis we used a new analysis procedure, 
%namely we employed linear splines to determine the unknown mass-over-charge ratios.
% Only interpolations have been used.

%Each 10~s Schottky frequency spectrum was analyzed independently.
Changes of the electron cooler voltage were done in steps of 0.5~kV
so that for adjacent cooler settings the measured frequency spectra have a significant overlap.
% Since several settings of the electron cooler have been recorded, there are many overlapping spectra. 
Furthermore, the same nuclide can be present in different charge states, which allows for a
redundant analysis. The nuclides whose masses have been measured
for the first time or which mass accuracy was improved in this work are listed in Table~\ref{newm}.

\begin{table}[t!]
\caption{Nuclides whose masses were determined for
the first time (in boldface) or whose mass uncertainty was improved in this work.
Listed are the proton ($Z$) and mass ($A$) numbers, the number of
experimental settings ($N_{set}$) in which this nuclide was
observed, and the obtained mass excess value ($ME$) with the
corresponding $1\sigma$ total ($\sqrt{\sigma^2_{stat}+\sigma^2_{syst}}$) uncertainty ($\sigma(ME)$).\label{newm}}
\begin{center}
\begin{tabular}{rrrrr}
\hline \hline
Z & A & $N_{set}$ &$ME$& $\sigma(ME)$ \\
   &     &                    & (keV) & (keV)                 \\
\hline
{\bf 71}&   {\bf 181}& {\bf 1} & {\bf -44797}& {\bf 126}\\
{\bf 71}&   {\bf 183}&  {\bf 1} & {\bf -39716}& {\bf 80}\\
\hline
{\bf 72}& {\bf 185}& {\bf 1} & {\bf -38320}& {\bf 64}\\
{\bf 72}& {\bf 186}& {\bf 1} & {\bf -36424}&{\bf 51}\\
\hline
{\bf 73}& {\bf 187}& {\bf 2} & {\bf -36896}&{\bf 56}\\
{\bf 73}&{\bf 188}& {\bf 2} & {\bf -33612}&{\bf 55}\\
\hline
74& 189& 5 &    -35618& 40\\
74& 190& 7 &    -34388& 41\\
{\bf 74}&{\bf 191}&{\bf 1} & {\bf -31176}&{\bf 42}\\
\hline
{\bf 75}&{\bf 192}&{\bf 1} & {\bf -31589}&{\bf 71}\\
{\bf 75}&{\bf 193}&{\bf 7} & {\bf -30232}&{\bf 39}\\
\hline
76& 195& 1 &    -29512& 56\\
\hline
\hline
\end{tabular}\end{center}
\end{table}

Since the calibration curve is not known exactly and is approximated with linear splines, and
since the number of calibration points in each spectrum is small,
there is inevitably a systematic error introduced by the analysis method.
Ways to estimate systematic uncertainty have been described in our previous works~\cite{Ra-NPA,Li-NPA,Chen2012}.
For this purpose in the present work, we re-determined the mass of each reference nuclide. 
This was done consecutively by setting each of the references as ``no''-reference and
obtaining its mass from the remaining 17 references. 
The re-determined mass excess values are listed in Table~\ref{references} along with their literature values~\cite{AME}. 
The systematic error $\sigma_{syst}$ has been obtained from solving the following equation:
\begin{equation}
\sum_{i=1}^{N_{ref}}\frac{(ME^{(i)}_{AME}-ME^{(i)})^2}{\sigma_{AME(i)}^2+\sigma_{(i)}^2+\sigma_{syst}^2}=N_{ref},
\end{equation}
where $N_{ref}=18$ is the number of reference nuclides, $ME^{(i)}$ ($\sigma_{(i)}$) 
and $ME_{AME}^{(i)}$ ($\sigma_{AME(i)}$) are the re-calculated and literature mass excess
values (statistical uncertainties) of the $i$-th reference nuclide, respectively.
The systematic uncertainty of the present analysis amounts to $\sigma_{syst}=38 $~keV. 
%By adding quadratically this uncertainty to the re-calculated mass values in Table~\ref{references}, 
%we obtain that for one nuclide,  $^{189}$Re, the mass deviates by 2.1$\sigma$, which is allowed for $N_{ref}=18$.
The final uncertainties listed in Table~\ref{newm} were obtained from a quadratic sum of the systematic and statistical uncertainties.
%: $\sigma(ME)_{(i)}=\sqrt{\sigma_{(i)}^2+\sigma_{syst}^2}$.

We note, that in contrast to Ref.~\cite{Chen2012} we do not observe any significant systematic dependence 
of the re-calculated mass values versus their proton number and correspondingly do not reduce the systematic errors.
A dedicated study should be performed to investigate the origin of this inconsistency.

\section{Discussion}

\begin{figure*}
\includegraphics[height=5.4cm]{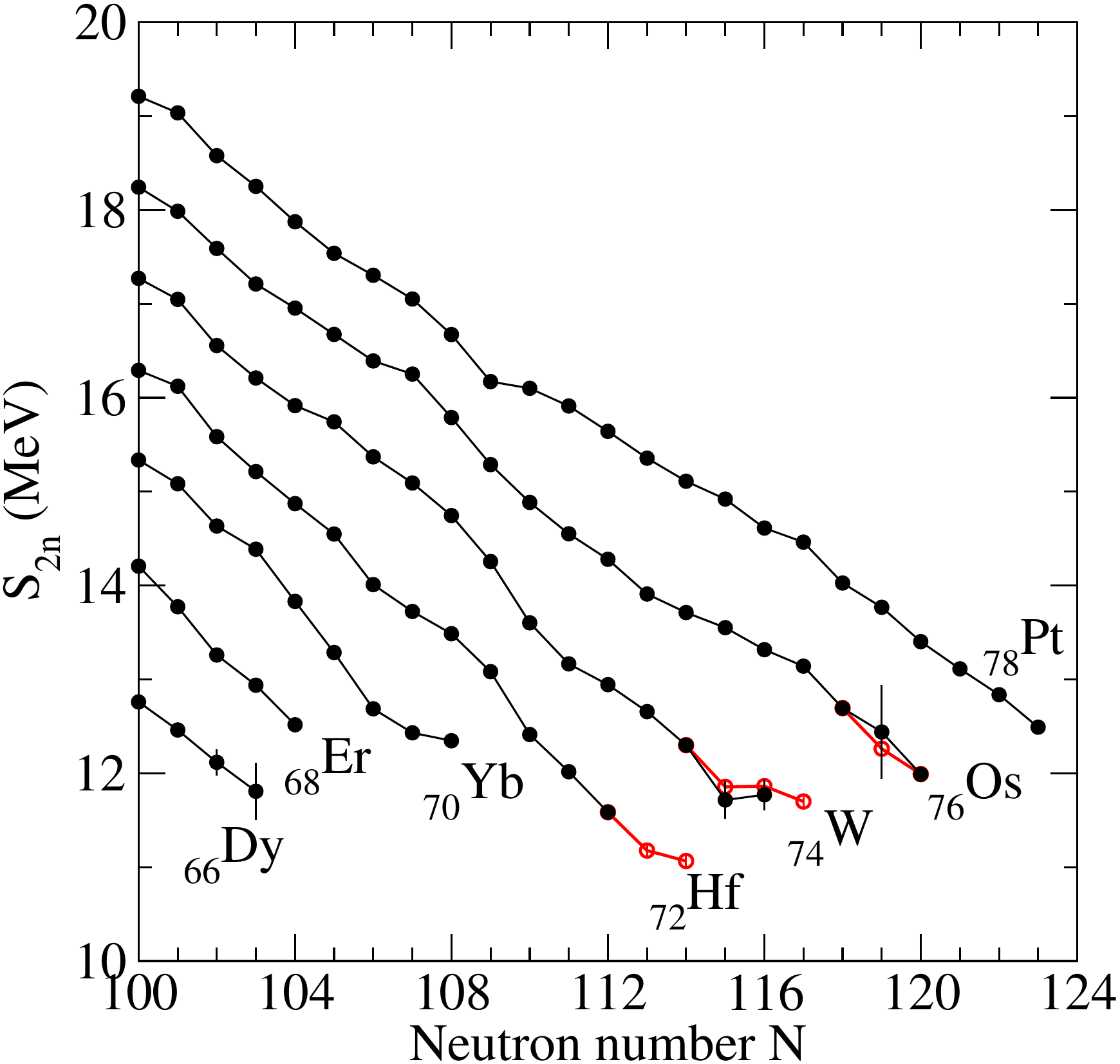}\hspace{8mm}
\includegraphics[height=5.3cm]{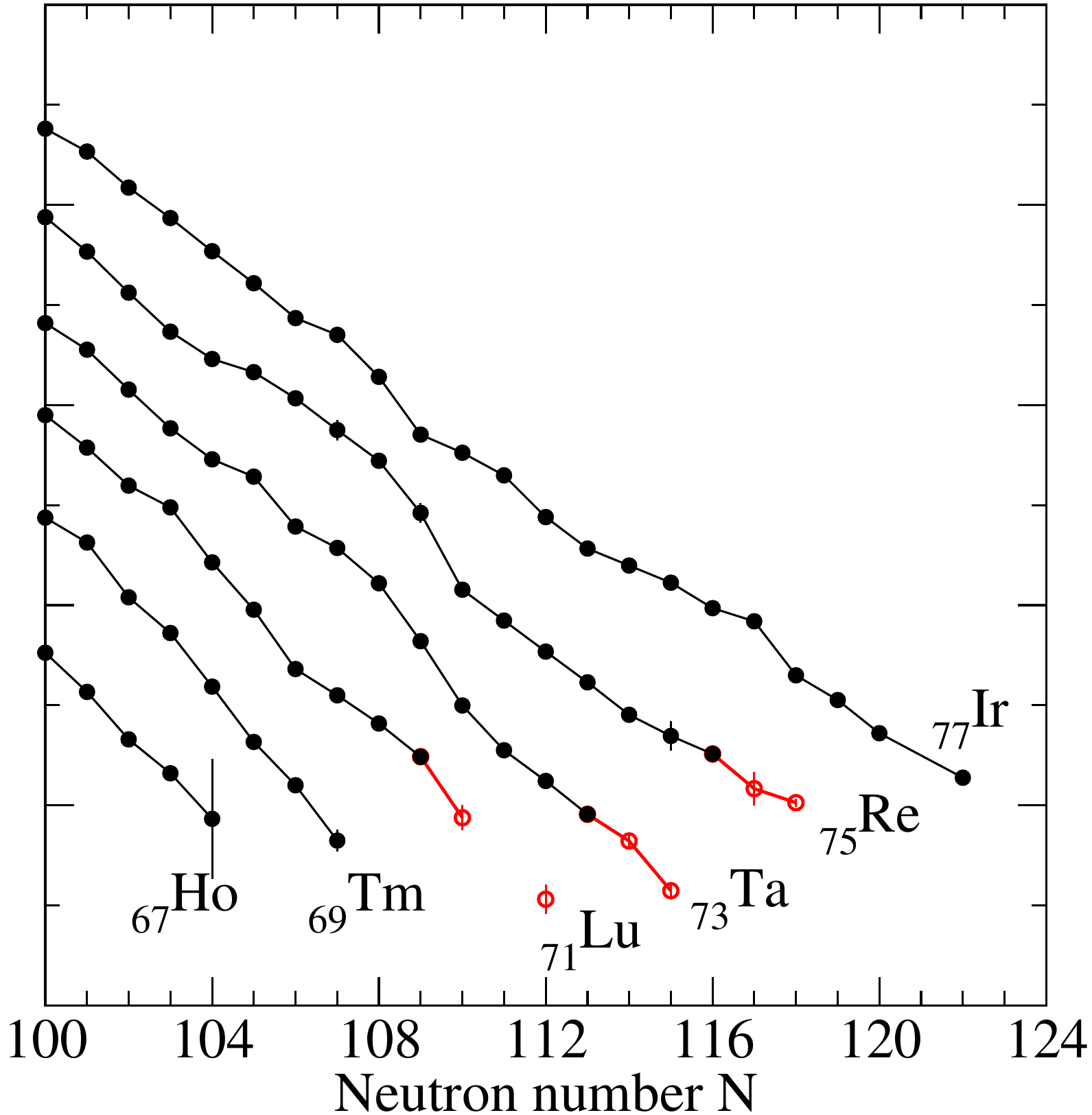}
\includegraphics[height=5.3cm]{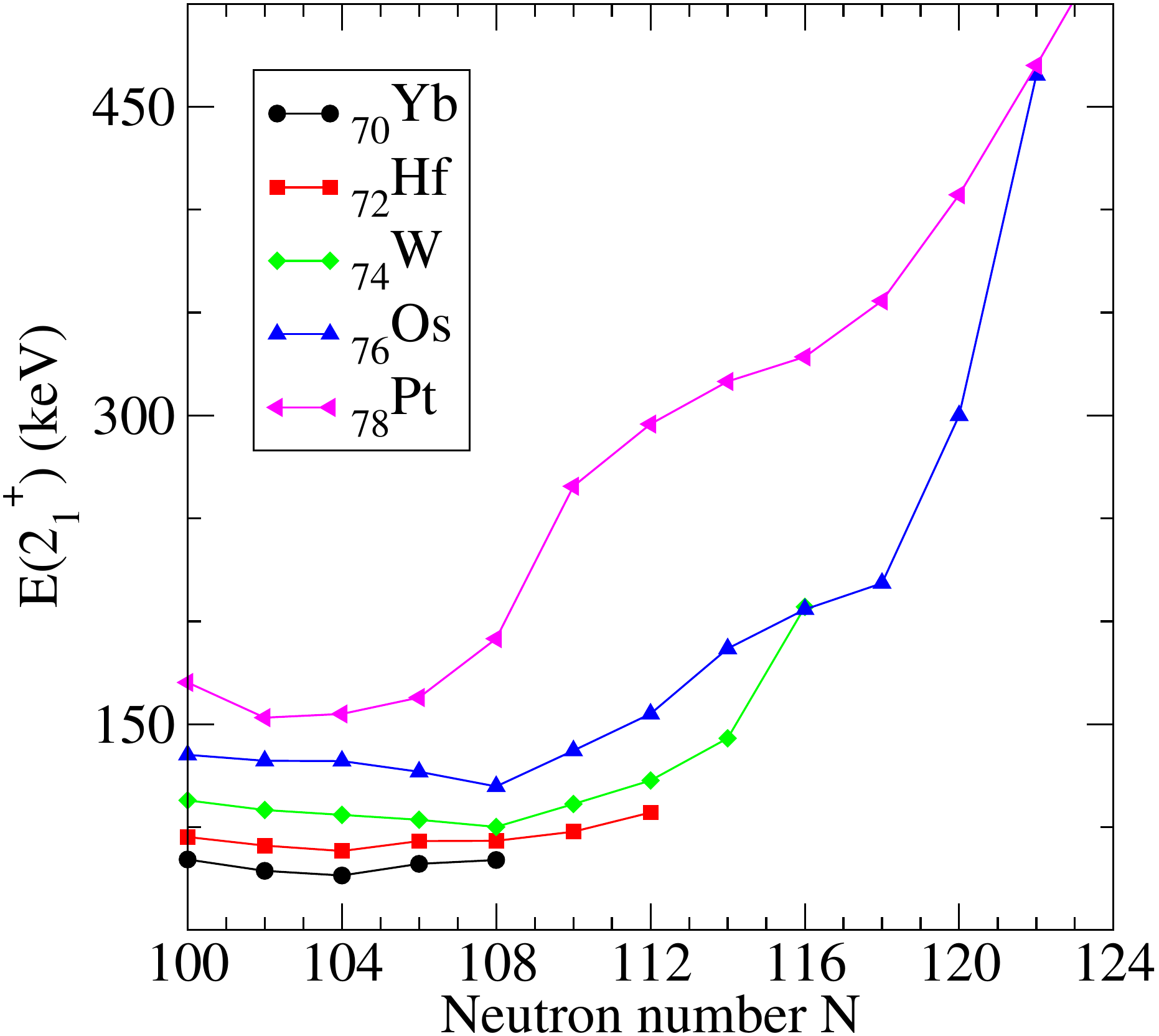}
\caption{(Color online) Data for the $A\sim180$ region,
two-neutron separation energies \cite{AME} as a function of
neutron number from $N=100$ to $N=124$ for even-$Z$ (left) from Dy
to Pt and odd-$Z$ (middle) from Ho to Re. The new $S_{2n}$ values
obtained from this work are shown in red color while the
literature $S_{2n}$ values are shown in black color. Right :
Energies of the first excited 2$^+$ states \cite{NDC} for the same
even-even nuclei as in the left panel.\label{s2n_e2}}
\end{figure*}

The new masses allow us to obtain interesting information on
nuclear structure. Fig.~\ref{s2n_e2} (left, middle) shows
two-neutron separation energies ($S_{2n}$) as a function of
neutron number for $Z=66-78$ in the $A\sim180$ region.
Fig.~\ref{s2n_e2} (left) is for even proton numbers while
Fig.~\ref{s2n_e2} (middle) is for odd proton numbers. The new
$S_{2n}$ values, for $^{181,183}$Lu, $^{185,186}$Hf,
$^{187,188}$Ta, $^{191}$W, $^{192,193}$Re, calculated from the
masses measured in this study, are marked in red color. For known
masses whose values were improved in this experiment,
$^{189,190}$W and $^{195}$Os, the literature $S_{2n}$ values are
illustrated with black color.

By inspecting Fig.~\ref{s2n_e2} (left), one can notice, that in
the $S_{2n}$ values of the even-$Z$ nuclei a flattening in Yb, Hf
and W is seen at almost the last neutron numbers experimentally
known. The flattening in $S_{2n}$(W), using the improved W masses,
is confirmed and $S_{2n}$ at $N=117$ continues with the same
behavior. In contrast to Hf and W, the new $S_{2n}$(Os) point,
which is a bit lower at $N=119$ than in previous measurements,
shows no flattening.

Fig.~\ref{s2n_e2} (middle) with the new measured masses does not
show similar effects in $S_{2n}$ as in Fig.~\ref{s2n_e2} (left).
However, there is a small change in slope (a more rapid fall-off)
at $N=110$ compared to the lower-$N$ trend in $S_{2n}$($_{71}$Lu),
$S_{2n}$($_{73}$Ta) and $S_{2n}$($_{75}$Re) (we also see this drop
for $_{72}$Hf and $_{74}$W in Fig.~\ref{s2n_e2} (left)), and maybe
at $N=115$ in $S_{2n}$($_{73}$Ta) as well. It is highly desirable
to have more odd-$Z$ mass measurements in this region for a
comparison with even-$Z$ where more data are available. Thus, we
concentrate below on discussing even-$Z$ nuclei.

Fig.~\ref{s2n_e2} (right) shows the energy of the first excited
2$^+$ states against neutron number for $Z=70-78$ in the
$A\sim180$ region. This important, simple, observable has a high
energy (can be a few MeV) at magic numbers where nuclei are
spherical and very low energies (less than 100 keV for well
deformed heavy nuclei) near mid-shell. The $E$(2$_1^+$) values
usually decrease smoothly between the beginning and middle of a
shell except when there is a sudden change in structure. Since
$N\sim104$ is mid-shell for the nuclei illustrated in
Fig.~\ref{s2n_e2}, $E$(2$_1^+$) has a minimum at or close to
$N=104$. After the mid-shell, the energy increases towards the
$N=126$ magic number.

Let us now focus on the W-Pt nuclei in Fig.~\ref{s2n_e2} and
compare the behavior of $E$(2$_1^+$) and $S_{2n}$. In particular,
we look at $S_{2n}$ in isotopes where $E$(2$_1^+$) changes
rapidly, indicating a sudden change in structure. In W,
$E$(2$_1^+$) increases from $N=114$ to 116 by a considerably
larger amount compared to the other W isotopes (see
Ref.~\cite{podolyak2000}). This jump signals a structural change
from approximately constant deformation to decreasing deformation
at $N=116$ (after $N=114$). Note that this neutron number is
exactly where $S_{2n}$ exhibits flattening.

Similar to W, Os at $N=120$ (after $N=118$) has a jump in
$E$(2$_1^+$). However, $S_{2n}$($^{196}$Os) does not reveal an
obvious change at the same neutron number. At first glance, this
seems inconsistent with the interpretation explained for W above
but, in fact, there might be an explanation for this different
behavior which would provide additional insight into the relation
of binding to collectivity.

\begin{figure}
\includegraphics[width=\linewidth]{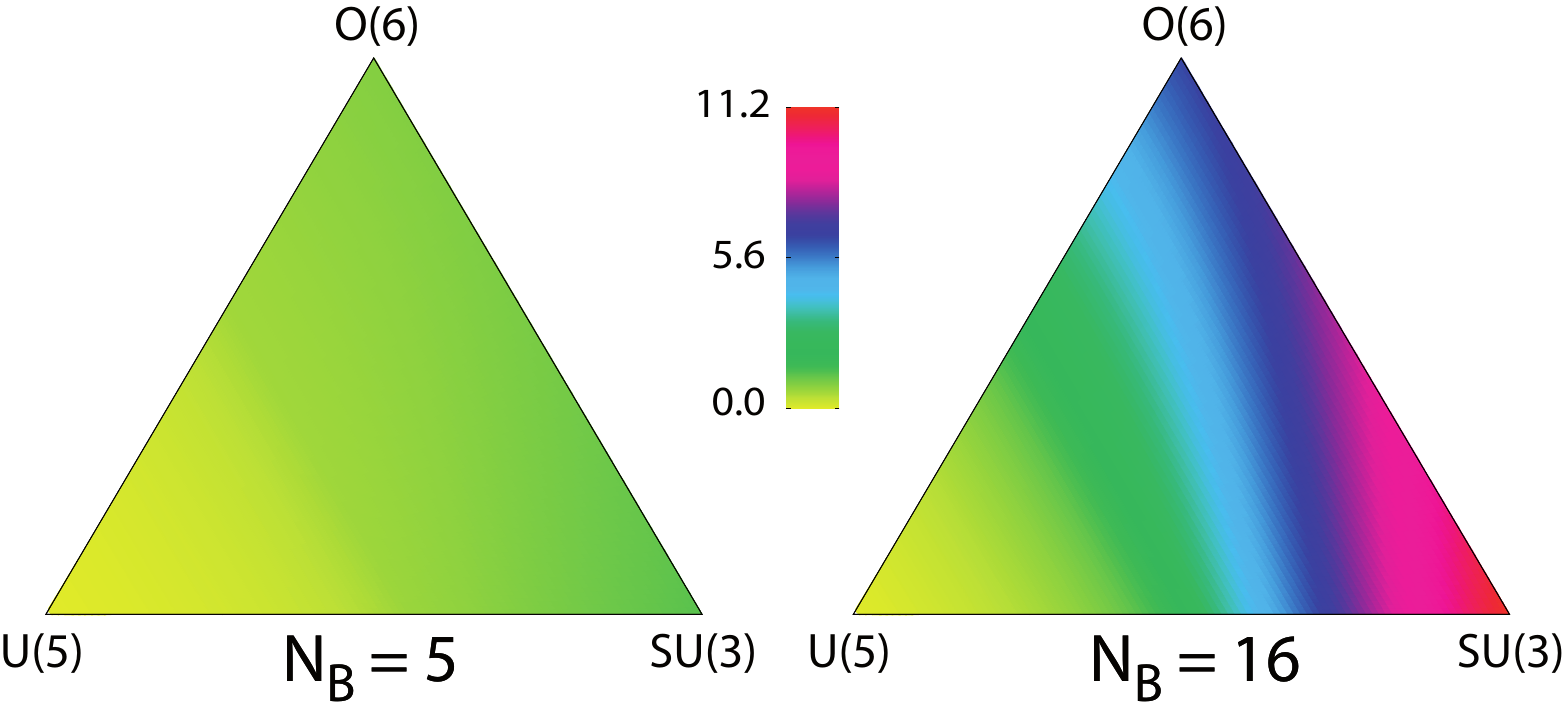}
\caption{(Color online) The symmetry triangle of the IBA showing
the three dynamical symmetries at the vertices. The colors
indicate calculated collective contributions in MeV to binding
energies for $N_B = 5$ (left) and $N_B=16$ (right). A similar
triangle for $N_B=16$ was presented in
Ref.~\cite{Cakirli2}.\label{triangle}}
\end{figure}

Reference~\cite{Cakirli2} showed the structural sensitivity of
calculated collective contributions to binding. In addition
Ref.~\cite{Cakirli2} stressed that collective binding is very
sensitive to the number of valence nucleons.
Calculated collective contributions to binding using the IBA-1
model for boson numbers $N_{B}=5$ (left) and $N_{B}=16$ (right)
are illustrated in the symmetry triangle \cite{Cakirli2} in
Fig.~\ref{triangle}. The three corners of the triangle describe
three dynamical symmetries, U(5) (vibrator), SU(3) (rotor) and
O(6) ($\gamma$-soft) (for more details, see Ref.~\cite{Iachello}).
The color code in Fig.~\ref{triangle} changes from yellow to red
when the collective effects increase. Needless to say, nuclei have
more valence particles (so boson numbers) around the SU(3) corner
than the U(5) (and also O(6)) corner. One sees that the collective
binding energy (B.E.) rapidly increases for nuclei with axial
deformation, that is, near SU(3). Note that the triangles are
presented for fixed boson numbers. In both, the color scale is
kept the same to point out that the collective B.E.s are larger in
N$_B$=16 than 5. As shown in Fig.~4 of Ref.~\cite{Cakirli2}, the
collective binding energies vary approximately as the square of
the number of valence nucleons in the context of IBA calculations.
Therefore, for a lower number of valence nucleons,
Fig.~\ref{triangle} shows similar trends for both  $N_{B}=5$ and
$N_{B}=16$, but the overall binding is considerably less (compare
(left) and (right) of Fig.~\ref{triangle}). We now suggest that
the behavior of $E(2_1^+)$ and $S_{2n}$ in Fig.~\ref{s2n_e2} can
be understood in terms of this dual dependence of binding on
collectivity and valence nucleon number.

If $^{190}$W and $^{196}$Os are mapped in the symmetry triangle,
$^{190}$W will likely be closer to the SU(3) corner than
$^{196}$Os \cite{Cakirli-pri}. One of the ways to understand this
is from the $P$-factor \cite{RCasten2}, defined as $P= N_p\cdot N_n / (N_p +
N_n)$, where $N_p$ denotes the number of valence protons (proton
holes) and $N_n$ the number of valence neutrons (neutron holes).
Thus $P$ is a quantity that can provide a guide to structure. For
example, the onset of deformation in heavy nuclei corresponds to
the $P$-factor around 4 and 5. Generally, if $P$ is larger than 3,
collective effects increase. That is, nuclei become deformed and
approach closer to the SU(3) corner.

\begin{figure}
\includegraphics[width=\linewidth]{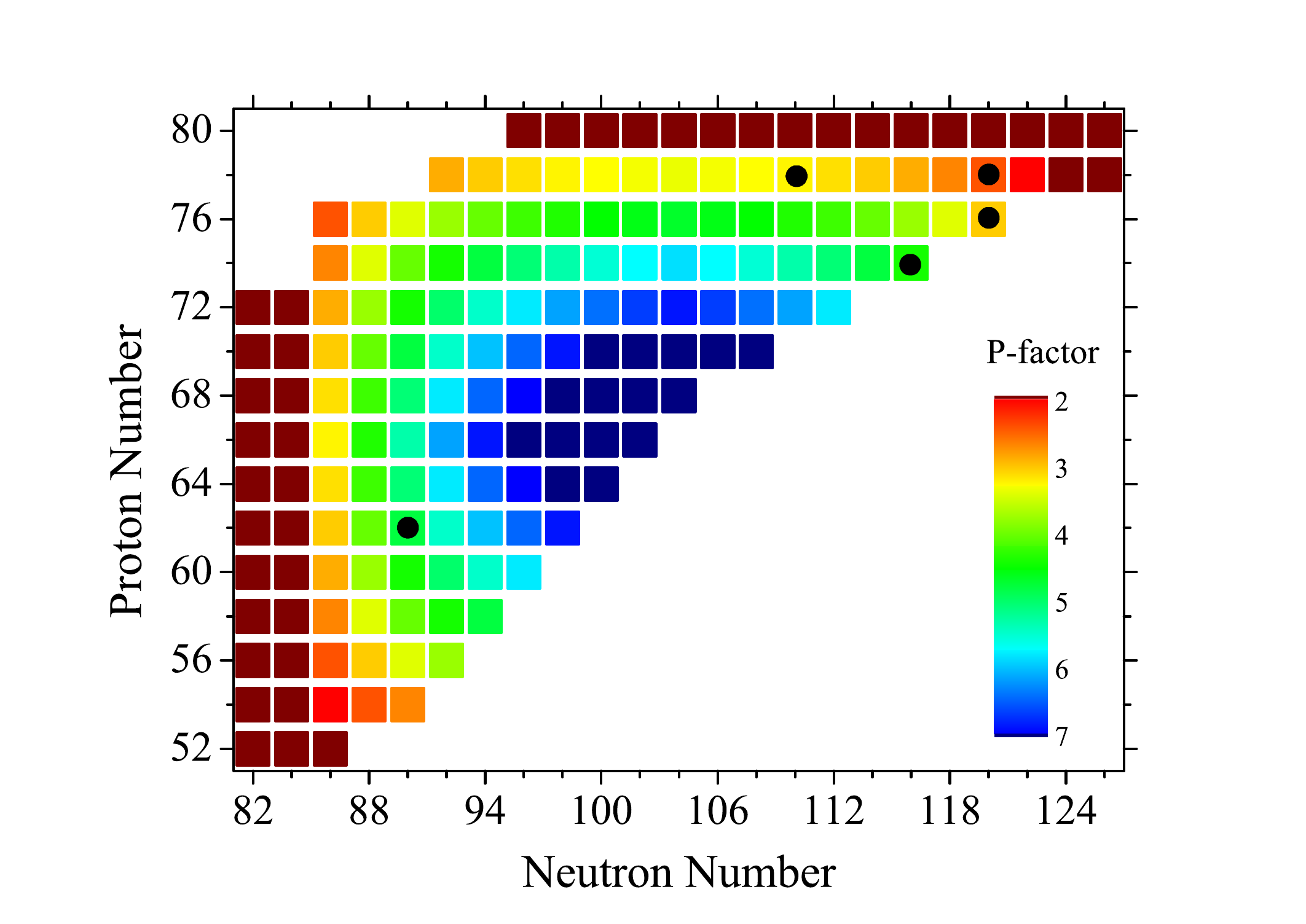}
\caption{(Color online) $P$-factor values illustrated with a color
code for even-even nuclei in the $Z=50-82$ and $N=82-126$ shells.
Black points marked are for the key nuclei discussed, namely,
$^{190}$W, $^{196}$Os, $^{188}$Pt,  $^{198}$Pt, and
$^{152}$Sm.\label{Pfactor}}
\end{figure}

Fig.~\ref{Pfactor} shows color-coded values for the $P$-factor
for the $Z=50-82$, $N=82-126$ region, and indicates the
$P$-factors for the nuclei relevant to this discussion. $^{190}$W
has 8 valence protons and 10 valence neutrons while $^{196}$Os has
$N_p=6$ and $N_n=6$. Correspondingly, these nuclei have
$P$-factors of 4.4 and 3, respectively. The greater collectivity
of $^{190}$W compared to $^{196}$Os suggested by their $P$-factors
is reflected in its lower 2$_1^+$ energies as seen in
Fig.~\ref{s2n_e2} (right). Thus, for two reasons -- both greater
collectivity and more valence nucleons -- the collective binding
should be much greater in $^{190}$W than in $^{196}$Os and changes
in binding energies ($S_{2n-coll}$ values) should be on a larger
scale. We suggest that this accounts for the fact that we see a
flattening in $^{190}$W clearly but not in $^{196}$Os in
Fig.~\ref{s2n_e2} (left). Obviously, it is very important to have
new data, both masses and spectroscopic information, on even more
neutron rich W isotopes although such experiments are difficult.
Even the mass of $^{192}$W alone would be telling since the trend
in $E$(2$_1^+$) is quite clear already.

\begin{figure}
\includegraphics[width=\linewidth]{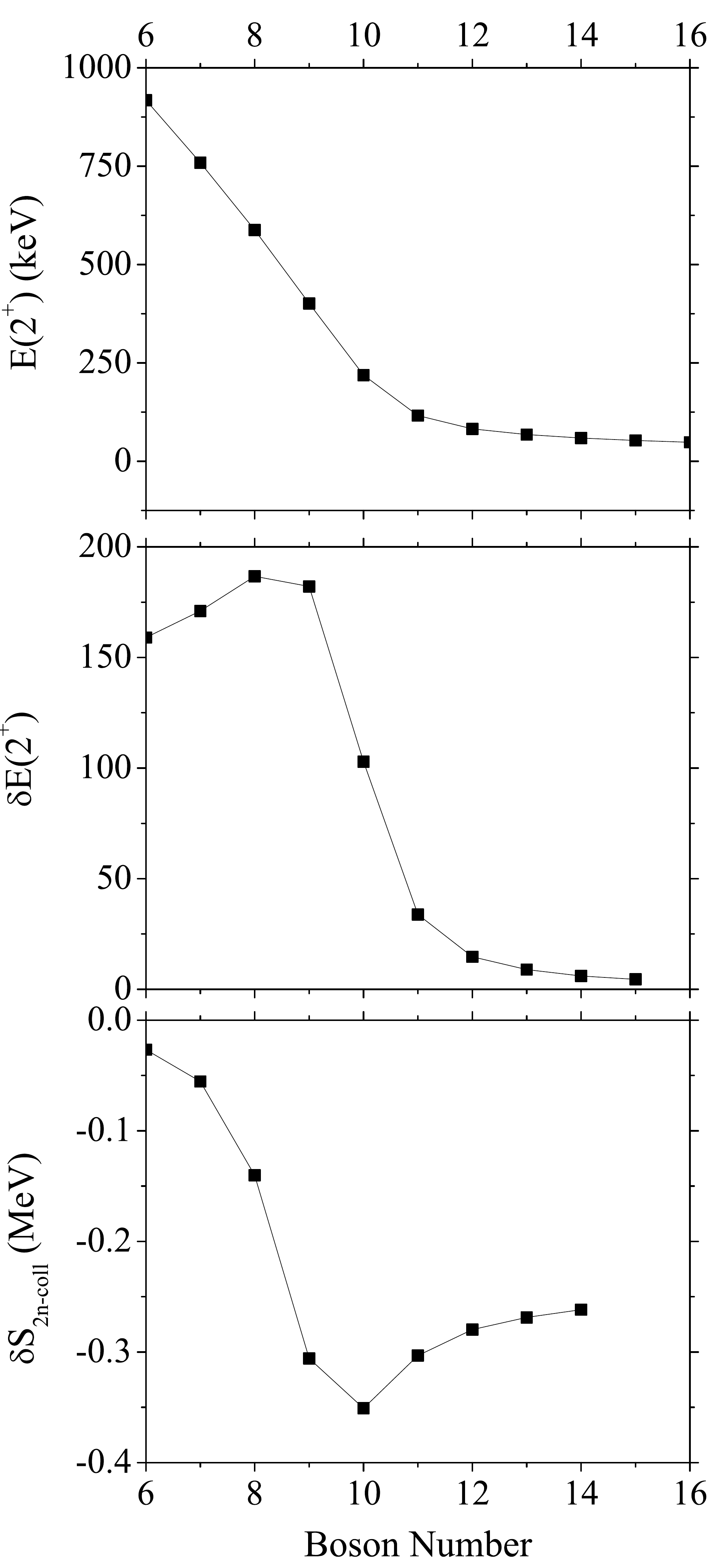}
\caption{Calculated $E$(2$_1^+$) (top), $\delta E$(2$_1^+$) (middle)
and $\delta S_{2n-coll}$ (bottom) values from IBA calculations as
a function of boson number $N_B$. The points for
$\delta E$(2$_1^+)$ and $\delta S_{2n-coll}$  correspond to a set
of schematic IBA calculations in which $\kappa$, and $\chi$ are
constant (at 0.02 and -1.32, respectively) while $\epsilon$, and
$N_B$ vary in a smooth way to simulate a spherical-to-deformed
transition region. The following equations are used for
$\delta E$(2$_1^+$) and $\delta S_{2n-coll}$:
$\delta E(2_1^+)(Z,N)=[E(2_1^+)(Z,N) - E(2_1^+)(Z,N+2)] /
E(2_1^+)(Z,N)$ and
$\delta S_{2n-coll}=-[S_{2n-coll}(Z,N) -
S_{2n-coll}(Z,N+2)$], respectively.\label{dS2n-E2}}
\end{figure}

Existing data on Pt nicely illustrate and support these ideas.
Fig.~\ref{s2n_e2} (right) shows two jumps in $E$(2$_1^+$) for
Pt, around $N\sim110$ and $N\sim118-120$. Looking at $S_{2n}$ for
Pt, there is a kink near $N\sim110$ but a smooth behavior near
$N\sim118$. For $^{188}$Pt, $N_p$ is 4 and $N_n$ is 16 so the
$P$-factor is 3.2. This isotope, with 20 valence nucleons, is
relatively collective and once again one sees an anomaly in
$S_{2n}$ as well. In contrast, for $^{198}$Pt$_{120}$ with only 10
valence nucleons, and a $P$-factor of only 2.4, the lower
collectivity (seen in the much higher 2$^+$ energy) and the lower
number of valence nucleons are such that $S_{2n}$ shows no
anomaly, but rather a nearly straight behavior.

This qualitative interpretation is supported by collective model
calculations. A thorough and detailed study of this or any
transition region requires a very careful and systematic
assessment of all the data on energies, transition rates, and
binding energies, the choice for the specific terms to include in
the Hamiltonian and the optimum approach to fitting the data. We
are undertaking such a study and will present the results in a
future publication \cite{Cakirli3}. Nevertheless, it is useful to
present an example of the model results here to validate the ideas
presented above. To this end, we have carried out a schematic set
of IBA calculations using the Hamiltonian \cite{7,8}

\begin{equation}
\label{eqH}
H = \epsilon \hat{n}_d - \kappa{Q} \cdot {Q}
\end{equation}

\noindent where $Q$ is a quadrupolar operator

\noindent

\begin{equation}
\label{eqQ} 
{Q}= (s^{\dagger}\tilde{d} +
d^{\dagger}s) + \chi(d^{\dagger}\tilde{d})^{(2)}.
\end{equation}

\noindent The first term in Eq.~\eqref{eqH} drives nuclei spherical while
the $Q\cdot Q$ term induces collectivity and deformation. Therefore a
spherical-deformed transition region involves a systematic change
in the ratio of $\epsilon$ to $\kappa$. No generality is lost by
keeping $\kappa$ constant (at 0.02 MeV). We follow a trajectory
along the bottom axis of the triangle corresponding to $\chi
=-1.3228$. Fig.~\ref{dS2n-E2} illustrates the results, for $N_B
= 6-16$ showing $E$(2$_1^+$) and the differentials of $E$(2$_1^+$) (for
$N_B = 6-15$) and for the collective contributions to $S_{2n}$,
$S_{2n-coll}$, (for $N_B = 6-14$). There is a clear change in
structure  at $N_B\sim$10 which is seen in a change in trend of
$E$(2$_1^+$). Between $N_B = 10$ and 11, R$_{4/2}$ changes from 2.60
to 3.13. This corresponds to a maximum in the normalized
differential of $E$(2$_1^+$). Confirming our association of
structural changes with kinks in $S_{2n}$, the differential of the
collective part of $S_{2n}$ also shows an extremum at exactly the
same point. These ideas will be expanded in our future publication
\cite{Cakirli3}.

\begin{figure}
\includegraphics[width=\linewidth]{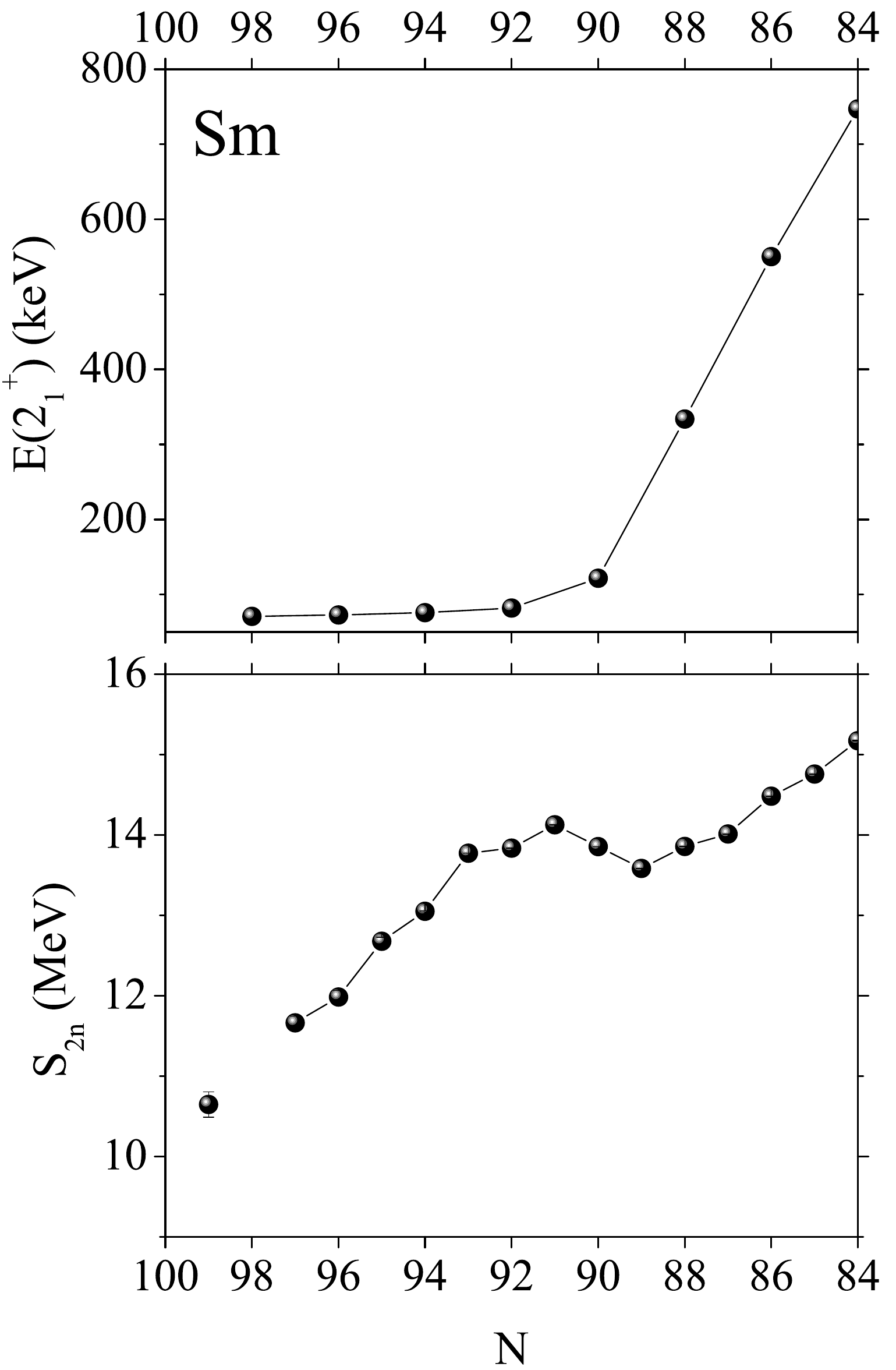}
\caption{Experimental $E$(2$_1^+$) (top) and
$S_{2n}$ (bottom) values against neutron number for $_{62}$Sm
\cite{AME11, NDC}. \label{Sm-S2n-E2}}
\end{figure}

Besides the experimental examples of a correlation of $E$(2$_1^+$) energies 
and $S_{2n}$ values discussed in the context of Fig.~\ref{s2n_e2},  
our interpretation can
easily be illustrated with the Sm isotopes around $N=90$. As is well
known, there is a sudden onset of deformation for the rare earth
nuclei from $N=88$ to 90. This effect is clear from various
observables. One example is seen in Fig.~\ref{Sm-S2n-E2} which
shows the experimental $E$(2$_1^+$) energies (top) and $S_{2n}$
(bottom) as a function of neutron number. Note that we plot these
against decreasing neutron number so that the deformed nuclei are
on the left and spherical ones on the right to make the comparison
with Fig.~\ref{s2n_e2} easier. Note also that the overall trend in
$S_{2n}$ is opposite from that in Fig.~\ref{s2n_e2} since $S_{2n}$
values decrease with increasing neutron number (going to the left in Fig.~\ref{Sm-S2n-E2} (bottom) which simply 
reflects the filling of the shell model orbits. The noticeable
deviation occurs near $N\sim90$ where there is
a distinct flattening. To correlate the trends in these two
observables in the $N=90$ region, one can use the same
interpretation as above for W at $N=116$, namely, if there is a
visible change at neutron number $N$ in $E$(2$_1^+$) and there are
many valence nucleons, we expect to see a change in the behavior
of $S_{2n}$. In Fig.~\ref{Sm-S2n-E2} (top), the $E$(2$_1^+$) change
occurs at $N\sim90$. The isotope $^{62}$Sm at $N=90$ has $N_p=12$ and
$N_n=8$ so it has 10 bosons and its $P$-factor is 4.8 (see
Fig.~\ref{Pfactor}). One therefore expects to see a change in
$S_{2n}$. Fig.~\ref{Sm-S2n-E2} (bottom) confirms this
expectation. The clear structural change at $N=90$ shown in
$E$(2$_1^+$) is correlated with a larger binding compared to the
general trend in $S_{2n}$ as a function of $N$.

Similar correlations can be seen in some other nuclei as well.
Further details will be discussed in Ref.~\cite{Cakirli3}.
However, here, it is worth mentioning two more examples marked in
Fig.~\ref{s2n_e2}. The case of Yb-isotopes is interesting. Yb at
$N=107$ starts to change slope in $S_{2n}$ and a flattening occurs
at $N=108$. The $P$-factor is $\sim7$. With the interpretation
above, one would expect to see a change in $E$(2$_1^+$) after
$N=106$, at $N=108$. However, there is no sudden change in
$E$(2$_1^+$) in $^{178}$Yb. To understand Yb around $N\sim108$
better, it might therefore be useful to have additional $S_{2n}$
values (mass measurements) and also more spectroscopic results for
the neutron-rich Yb isotopes.

Hf at $N=114$ has a $P$-factor 5.4 and one sees a flattening in
$S_{2n}$. The corresponding 2$^+$ energies, however, are not
known. Thus, similarly as in Yb, we need more spectroscopic
results for Hf.

To summarize, we observed a correlation between the behavior of
$S_{2n}$ obtained from our measured masses with the spectroscopic
data for $E(2_1^+)$, which could be related to nuclear
collectivity and valence nucleon number.

\section{Conclusion}

Direct mass measurements of neutron-rich $^{197}$Au projectile
fragments at the cooler-storage ring ESR yielded new mass data.
Masses of nine nuclides were obtained for the first time and for
three nuclei the mass uncertainty was improved.

With the new masses, two-neutron separation energies, $S_{2n}$,
are investigated. We showed that changes in structure, as
indicated by changes in the collective observable $E$(2$^+_1$),
are reflected in $S_{2n}$ values in nuclei such as $^{190}$W and
$^{188}$Pt, which have large $P$-factors, are collective,  and
have large valence nucleon numbers. For nuclei with similar
changes in $E$(2$_1^+$), such as $^{196}$Os, and $^{198}$Pt, which
have lower collectivity and $P$-factors, and fewer valence
nucleons, the sensitivity of collective binding to structure is
greatly reduced and smooth trends in $S_{2n}$ are observed. In Hf,
there are new $S_{2n}$ values at $N=113$, 114 where we see a
flattening but there is no spectroscopic data at $N=114$. To
confirm the ideas discussed in this paper and also in
Ref.~\cite{Cakirli3}, it would be useful to measure the
$E$(2$_1^+$) for Hf at $N=114$. Similarly, mass and spectroscopic
measurements are suggested for nuclei such as Yb with $N\sim108$.
To conclude, these new data illustrate subtle changes in structure
and the correlation with $E$(2$_1^+$) reveals a valuable way to
correlate changes in structure in terms of both masses and
spectroscopic observables. Of course, to quantitatively test these
ideas requires a systematic collective model study of the
mass-structure relationship in this region. Such a project has
been initiated~\cite{Cakirli3} and we illustrated some of the
results here.

\section{Acknowledgments}

The authors would like to thank the GSI accelerator team for the excellent technical support. 
%Fruitful discussions with K. Blaum, S.Yu. Torilov, X. Ma, H. Xu, X. Tu and Y. Yuan are gratefully acknowledged. 
This work was supported by the BMBF Grant in the framework of the Internationale Zusammenarbeit in Bildung und Forschung Projekt-FKZ 01DO12012,
by the Alliance Program of the Helmholtz Association (HA216/EMMI), by the Max-Planck Society and the US DOE under Grant No. DE-FG02-91ER-40609.
D.S. is supported by the International Max Planck Research School for Precision Tests of Fundamental Symmetries at MPIK. 
R.B.C. thanks the Humboldt Foundation for support.
K.B. and Y.A.L. thank ESF for support within the EuroGENESIS program.
K.B. acknowledge support by the Nuclear Astrophysics Virtual Institute (NAVI) of the Helmholtz Association. 
Z.P. would like to acknowledge the financial support by Narodowe Centrum Nauki (Poland) grant No. 2011/01/B/ST2/05131. 
M.S.S. acknowledges the support by the Helmholtz International Centre for FAIR within the framework of the LOEWE program launched by the State of Hesse. 
B.S. is partially supported by NCET, NSFC (Grants No. 10975008, 11105010 and 11035007).
P.M.W. acknowledges the support by the UK STFC and AWE plc.
T.Y. is grateful for a grant-in-aid for a scientific research No. A19204023 by the Japanese Ministry of Education, Science and Culture.

\end{document}